\documentclass[11pt]{article}
\usepackage{setspace}

\def\be{\begin{equation}}

\def\ee{\end{equation}}

\def\bea{\begin{eqnarray}}

\def\eea{\end{eqnarray}}

\newcommand\eps{\epsilon}
\newcommand\refeq[1]{(\ref{#1})}
\begin{document}

\singlespace

\begin{flushright} BRX TH-623 \\
CALT 68-2798
\end{flushright}

\vspace*{.3in}

\begin{center}

{\Large\bf  Is BTZ a separate superselection sector of CTMG? }

{\large S.\ Deser}

{\it Physics Department,  Brandeis University, Waltham, MA 02454 and \\
Lauritsen Laboratory, California Institute of Technology, Pasadena, CA 91125 \\
{\tt deser@brandeis.edu}
}

{\large J.\ Franklin}

{\it  Reed College, Portland, OR 97202 \\
{\tt jfrankli@reed.edu}}

\end{center}

\begin{abstract}
      We exhibit exact solutions of (positive) matter coupled to original ``wrong $G$-sign" cosmological TMG. They all evolve to conical singularity, rather than to black hole -- here negative mass -- BTZ. This provides evidence that the latter constitute a separate ``superselection" sector, one that unlike in GR, is not reachable by  physical sources.
\end{abstract}

\section{Introduction}

  \hskip .4in Topologically massive gravity's [1] cosmological, AdS, 
version (CTMG) [2] has recently become extremely popular, particularly in connection with 
its boundary excitations at the ``chiral point" -- where the dimensionless product of 
its mass parameter $\mu$ in~\refeq{FEOM} times the cosmological length 
$\ell \equiv (-\Lambda)^{-1/2}$ is unity [3]. Central in these discussions is the 
gravitational constant's sign: On the one hand, in the original TMG, that sign 
must be ``wrong" -- opposite to that of $G$ in $D=4$ GR -- in order that TMG's 
bulk excitations have positive energy [1]. On the other, at the chiral point, these modes coexist [4,5] with surface excitations that seem to require [3] the ``right" 
sign of $G$ to avoid instability, as do the central charges of the dual CFT [3,6]. The same  
sign conflict affects the black hole (BH) BTZ [7] of cosmological, (AdS), 
GR [8], whose solutions automatically also satisfy (either $G$-sign) CTMG.  But for our wrong sign, the BH becomes a negative energy state, potentially upsetting the model's 
stability.   At least this, BTZ, part of the sign dilemma can be dehorned if, 
as conjectured in [4], the BH BTZ lies in a different super-selection sector from normal, positive, 
matter coupled to wrong sign CTMG: If it cannot be reached physically by time-evolution 
of these systems, BTZ poses no stability risk.

    The above somewhat involved introduction explains the present work's 
motivation: we will exhibit examples of just this super-selection mechanism, 
namely exact solutions of full wrong sign CTMG whose physical, positive, matter 
distribution's endpoints are indeed the (conically) singular, rather than the BH BTZ-branch.  While examples can of course only provide evidence for, rather than prove, super-selection, the present favorable evidence is reassuring.

 Since bounded exact solutions of even the original 
($\Lambda = 0$) source-free TMG have never previously been found, it may seem 
miraculous that any can be produced in this more complicated sector.  The miracle hinges on an {\it a priori} surprising general theorem 
originally found for source-free TMG [9], extended to allow sources [10], stating that circularly 
symmetric, null, $T^\mu_\mu = 0$, matter+CTMG reduces to cosmological $D=3$ GR. 
This result enables us to study the much simpler GR system, for which null solutions with 
the usual ``right" $G$ sign are known.  Indeed we will simply flip their sign, or equivalently 
that of $T_{\mu\nu}$ to obtain ours. This will convert those spaces' original BH BTZ 
end-state to its conical singularity branch, which in turn means that these collapsing matter 
systems lie in a different super-selection sector from BH BTZ\footnote{A related result, from the 
other end, states that BTZ is stable under small perturbations [11], so it does not spontaneously 
decay into states with matter to spoil super-selection. }.

\section{CTMG and its sources}

  \hskip .4in  We begin with a lightning review of CTMG and of the 
``decoupling" theorem for circular symmetry that makes explicit 
solutions possible. Its gravitational part is the sum of the (AdS) Einstein 
and Cotton tensors,
\begin{equation}\label{FEOM}
 \left[G^{\mu\nu} +\Lambda\, g^{\mu\nu}\right] + 1/(\mu\, \sqrt{-g}) \, 
\eps^{\mu\alpha\beta}\,  D_\alpha (R^\nu_{\ \beta} - 1/4  \delta^\nu_{\ 
\beta}  R) = G\, T^{\mu\nu}.
\end{equation} 
The density $\eps^{\mu\alpha\beta}$ is of course $\eps^{0ij}$; but for circular 
symmetry there can be no $\eps^{0ij}$ in any of the other, ``even", terms: hence 
the Cotton tensor must vanish by itself, leaving pure GR coupled to matter. However, 
because Cotton is the conformal curvature tensor in $D=3$, its vanishing must still be
taken into account: it requires that space-time be conformally flat, imposing the condition 
on the GR sector that the source's $T^\mu_\mu$ vanish -- matter must be null. This 
defines the arena of our $D=3$ systems: null circularly symmetric matter coupled to AdS 
($\Lambda < 0$) GR. Fortunately, there exist several exact ``right" sign $G$ solutions of just
this type; the (trivial to execute) effect of flipping $G$ there changes the endpoint of 
their evolution from black hole, to conically singular BTZ:  For these systems then, BTZ lies in a 
separate ``super-selection" sector, as we will show.

     We will consider several different matter sources, noting how their $G$-flipped versions evolve to singular BTZ. [Incidentally, since there are no gravitational forces in $D=3$, there is no ``braking" of the matter infall process when $G$ flips.] Our first example is a null source generating the ($D=3$) Vaidya metric [12]; it will then be extended to other null systems [13]. 
We also treat the ($D=3$ version of an originally $D=4$) null infalling circular mass distribution [14], 
and briefly mention some non-null solutions [15] that indirectly buttress our case.

  \section{Explicit Solutions}

 \hskip .4in Our first example is the $D=3$ AdS GR version of the 
Vaidya metric,
\begin{equation}\label{Vstart}
ds^2 = -f(r,v) \, dv^2 + 2 \, dr \, dv + r^2 \, d\theta^2
\end{equation}
where $v$ is advanced time and the stress tensor's only non-vanishing 
component is
\begin{equation}
T_{vv} = \rho(v)/4 \pi  r , \, \, \, \, \, \, \, T^\mu_{\, \mu} = 0.
\end{equation}
This metric is determined by $\rho$ according to:
\begin{equation}\label{frv}
f(r,v)= \ell^{-2}\,  r^2 - g(v)  , \, \, \, \, \, \, \, g(v) 
\equiv\int^v \rho(v) \, dv.
\end{equation}

For zero source, $g(v)$ is just an integration constant $M$ (of either 
sign!), namely the BTZ solution.   Being the integral of $G \rho$ ($G$ 
is of course implicit in $T_{vv}$), $g(v)$ becomes negative for our flipped 
$G$, since we keep $\rho>0$: by~\refeq{frv}, this manifestly evolves to 
$M< 0$, singular, BTZ.  For non-zero $\rho$, as discussed in [12], there is an apparent horizon 
determined by $g(v=\infty)$; if its radius $R$ is not positive, then there is 
no horizon. But since $g(v)$ is the 
integral of $\rho$, it is negative for our flipped $G$ (there is of course an implicit 
$G$ in $T_{vv}$), and we are indeed on the negative $M$ singular branch of BTZ, 
as claimed.  [We emphasize that the {\it parameter} $M$ in BTZ corresponds
to {\it physical mass} $-M$ in wrong sign $G$, but that the metric's 
BH/singular nature is still decided by the positive/negative sign of its $M$ parameter.]

The Vaidya metric is actually the special case $k=1$ of a whole class [13] of null fluid 
systems defined by the pressure/density ratio, $P/\rho=k$.  Another soluble value considered 
there is the pressureless one with $k=0$. There is also no apparent horizon $R$ there for
wrong sign $G$; we merely quote the value
\begin{equation}
R= \ell^2 B \pm \sqrt{\ell^4 \, B^2 - \ell^2 \, \left(1 - 2 \, A  \right)}
\end{equation}
where $(A,B)$ are, here negative, matter-related constants, so $R<0$: no 
horizon. Generic $k$ values lead to higher polynomial equations for $R$, 
but there is no indication that a horizon can ever appear there either.

 A  different example, adapted from the $D=4$ model of 
[14], is a null matter delta-function circular ring of surface density 
$\mu$, the coefficient of $\delta(r)/r$. It separates the exterior (+) 
and interior (-) regions of the metric
\begin{equation}
ds^2 = -f_\pm \, dt^2 + f_\pm^{-1} \, dr^2 + r^2 \, d\theta^2
\end{equation}
  where $f_\pm= A_\pm + (r/\ell)^2$, and $(A_+,A_-)= (-M,0)$ are 
respectively massive/massless BTZ. Matching across the shell's curvature 
discontinuity relates $\mu$ and $M$,  according to (in units of $G$)
\begin{equation}
   16\,  \pi \, \mu=A_- - A+ =M
\end{equation}
and as expected, the sign of $M$ mirrors that of $\mu$:  once again,``wrong $G$" 
means once again singular BTZ endpoint $M<0$.

 Our conclusions regarding infall to the singular branch 
of BTZ are of course not confined to null sources of ``wrong sign" AdS 
GR -- more generally, non-null positive source solutions exhibit similar 
evolution to negative mass BTZ, as can be seen, for example, upon 
changing $G$'s sign in various models given in [15]. One explicit case 
there illustrates the outcome: the BTZ mass parameter $M$ of a collapsing dust cloud is
\begin{equation}
M = A \, G \, \rho - 1
\end{equation}
where $A$ is a positive constant, $\rho$ the (positive) density and our 
$G$ is negative; hence evolution is again to the singular BTZ endpoint here.

\section{Discussion}

\hskip .4in Solving CTMG's otherwise intractable field equations to analyze BTZ's interactions with matter required using highly symmetric sources. The resulting narrower class of solutions' evolution to singular BTZ can only constitute encouraging indication, rather than proof, that BTZ lies in a separate super-selection sector. Should these indications prove correct, they will have removed one important aspect of the CTMG sign dilemma.

SD thanks S.Carlip for essential discussions at this project's inception, and acknowledges support from NSF PHY 07-57190 and DOE DE-FG02-164 92ER40701 grants.

%\vfill\eject

\end{document}